# NOISE IN INTENSE ELECTRON BUNCHES*

S. Nagaitsev[†,1], Y.-K. Kim, The University of Chicago, Chicago, IL 60637, USA
Z. Huang, G. Stupakov, SLAC, Menlo Park, CA 94025, USA
D. Broemmelsiek, J. Jarvis, A. H. Lumpkin J. Ruan, G. Saewert, R. Thurman-Keup, Fermilab, Batavia, IL 60510, USA

*Abstract*

We report on our investigations into density fluctuations in electron bunches. Noise and density fluctuations in relativistic electron bunches, accelerated in a linac, are of critical importance to various Coherent Electron Cooling (CEC) [1-5] concepts as well as to free-electron lasers (FELs). For CEC, the beam noise results in additional diffusion that counteracts cooling. In SASE FELs, a microwave instability starts from the initial noise in the beam and eventually leads to the beam microbunching yielding coherent radiation, and the initial noise in the FEL bandwidth plays a useful role. In seeded FELs, in contrast, such noise interferes with the seed signal, so that reducing noise at the initial seed wavelength would lower the seed laser power requirement [6-8]. Status of the project will be presented.

## INTRODUCTION

Noise and density fluctuations in relativistic electron bunches, accelerated in a linac, are of critical importance to various Coherent Electron Cooling (CEC) concepts as well as to free-electron lasers (FELs). For CEC, the beam noise results in additional diffusion in a cooled beam that counteracts cooling; and if this noise is not controlled at sufficiently low level, the noise heating effects can overcome cooling. There have been several proposals in the past to suppress the noise in the beam in the frequency range of interest in order to optimize the cooling effects. In SASE FELs a microwave instability starts from the initial noise in the beam and eventually leads to the beam microbunching yielding coherent radiation, and the initial noise in the FEL bandwidth plays a useful role. In seeded FELs, in contrast, such noise interferes with the seed signal, so that reducing noise at the initial seed wavelength would lower the seed laser power requirement.

Advanced cooling and FEL concepts not only require the knowledge of beam noise level but also call for its control. We are proposing to carry out a systematic theoretical and experimental study of electron beam noise at micrometer wavelengths at the Fermilab FAST facility [9, 10]. Figure 1 shows the energy kick, experienced by a proton in the EIC CEC kicker section [11]. The longitudinal scale of the wake is $\sim 3$ μm, corresponding to the frequency bandwidth of interest of $\sim 40$ THz. This wavelength-scale is of general interest in accelerator and beam physics as indicated by the community-driven research opportunities survey. Our research goals are (1) to measure the electron beam density noise level in a 0.5 – 10-μm wavelength range, (2) to predict the beam noise level in order to compare with the measurements, and (3) to find mechanisms that affect the beam noise to control its level in a predictable manner. In this paper we will describe our progress to date as well as our future experimental plans at Fermilab's FAST electron linac.

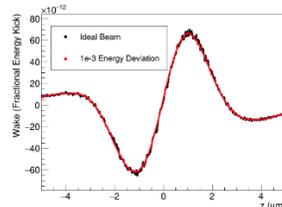

Figure 1: The energy kick, generated by one proton in the CEC kicker section. The longitudinal scale of this kick is $\sim 3$ μm, corresponding to the frequency bandwidth of interest of $\sim 40$ THz.

## FAST FACILITY AND APPARATUS

The Fermilab Accelerator Science and technology (FAST) Facility, shown in Fig. 2, is well-suited for this research as it can provide electron bunches with charges 0 – 3 nC, 1-30 ps long rms and energies 50 - 300 MeV. This makes it perfectly relevant to future needs of electron-ion colliders as well as injectors for future FELs. Electron bunches are generated by a $Cs_2Te$ photocathode and a UV laser. An L-band rf gun accelerates the beam to 4.5 MeV (typical). The facility also has a single-stage bunch compressor and a 100-m long FODO-based transport channel (not shown) allowing for several experimental stations. For example, Table 1 compares the FAST beam parameters with that of an EIC CEC concept.

Table 1: FAST and Proposed CEC beam parameters.

| Parameter | FAST | EIC | EIC |
|---|---|---|---|
| Proton beam energy, GeV |  | 100 | 275 |
| Elect. beam energy, MeV | 50 - 300 | 50 | 137 |
| Bunch charge, nC | 0 – 3 | 1 | 1 |
| Emittance, rms norm, μm | $\sim 3$ (at 1 nC) | 2.8 | 2.8 |
| Bunch length, mm | 0.3-10 | 12 | 8 |
| Drift section (amplifier), m | 80 | 100 | 100 |

___________
* Work supported by the DOE ARDAP office
† nsergei@fnal.gov
1 also at Fermilab

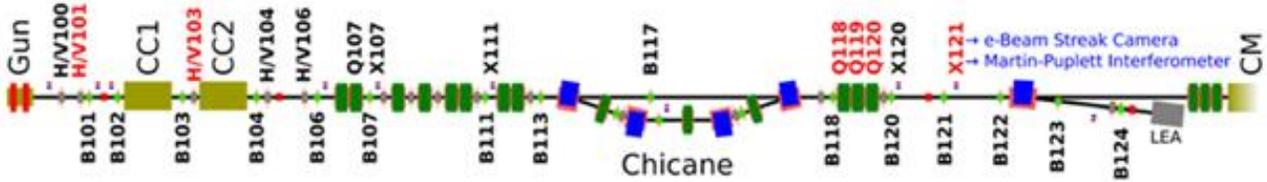

Figure 2: Diagnostics cross X121 (upstream of the SRF cryomodule) is presently equipped with an OTR screen, an Al-coated Si substrate, positioned at 45 degrees with respect to beam.

The initial principal experimental apparatus, Fig. 3, will consist of a thin highly reflective Al-coated Si wafer, inserted into the beam, in order to generate transition radiation (TR). Such a foil is already in use at FAST at the X121 location [12, 13]. The TR has a broad spectrum, extending from UV to tens of micrometers and containing the information about the beam density micro-structure. Also, the divergence of radiation has weak dependence on the wavelength.

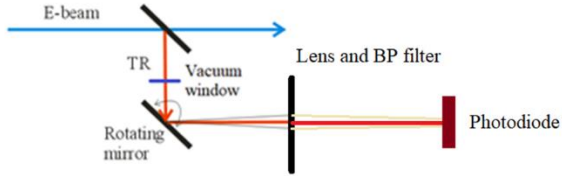

Figure 3: The proposed schematic of the TR spectrometer (BP is a band-pass filter).

Transition radiation energy, radiated spontaneously per unit frequency, $\omega$, per unit solid angle is:

$$\frac{d^2W}{d\omega d\Omega} = \frac{Z_0 q^2}{4\pi^3} \frac{\beta^2 \sin^2\theta}{\left(1-\beta^2 \cos^2\theta\right)^2}$$
$$\approx \frac{Z_0 q^2}{4\pi^3} \frac{\theta^2}{\left(\gamma^{-2}-\theta^2\right)^2}, \quad (1)$$

where $q$ is the bunch charge, $\beta = v/c$, and $Z_o \approx 377\ \Omega$ is the impedance of free space. After integrating Eq. (1) over the solid angle, we obtain the spectral radiation energy density. Limiting our collection angle to 90 degrees, we obtain the following energy per 1-nC bunch: 27 pJ (0.5 – 1 μm) and 21 pJ (1 – 5 μm). Thus, in the range of 0.9 – 2 μm we are expecting about 1 pJ per 100-nm bandwidth. The radiation energy Eq. 1 corresponds to a random electron distribution or the shot-noise level. Any excess above the shot-noise TR energy level, would have to be carefully considered as potential noise (or beam clumping) above shot noise. To this end, we will measure both the absolute value of TR energy and its dependence on bunch charge, where we expect a linear dependence, as in Eq. 1 for pure shot noise.

To measure the single-bunch TR energy, we plan to take advantage of the existing infrastructure in the FAST linac. The basic electron-beam diagnostics including rf BPMs, beam profile stations, and current monitors in the low-energy linac will be used. The X121 station (see Fig. 4) has a YAG:Ce screen viewed by a 5-Megapixel digital CCD camera for transverse beam size and emittance measurements (based on an upstream quadrupole field scan technique using Q118-120). The first beam noise experiments will be performed at 45 MeV using the X121 optical transition radiation (OTR) screen which is an Al-coated Si wafer oriented at 45 degrees to the beam direction. The X121 station's OTR can be transported by the all-Al-mirror optical transport as shown in Fig. 4 to either the InGaAs photo diode (with a 100-nm band-pass filter) and amplifier or to a Hamamatsu C5680 synchroscan streak camera, both located in an optical hut outside of the tunnel. The latter will provide ~2-ps (sigma) resolution for the 10- to 20-ps long (rms) bunches. A synchronous sum of a short micro-pulse train will be used to improve statistics for the streak camera images. The Al mirrors in the optical transport should be about 88% reflecting at 1 μm subject to oxidation effects. These 10 mirrors will be upgraded to Ag-coated mirrors that are 97% reflecting at the initial wavelengths of interest to reduce the losses in transport. Also, a dozen of 100-nm BP filters in the range of 0.9 – 2 μm were procured.

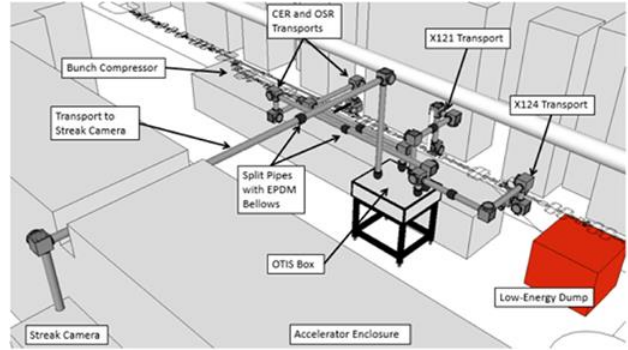

Figure 4: A schematic of the all-mirror optical transport from X121 to the streak camera and photodiode.

A schematic of the experimental setup is shown in Fig. 3. The experiments will use a single electron micropulse (bunch) with a digitizer used on the photodiode signal output. Fig. 5 shows the schematic of the photodiode detector.

Calibration of the photodiode / integrator was accomplished via two different laser systems: a 1054-nm Nd:YLF laser with a ~5-ps rms duration, and a 1550-nm Erbium fiber laser with a several-picosecond rms duration. The energy of the 1054 nm laser was measured with a Thorlabs S121C Si photodiode energy meter, and the energy of the 1550 nm laser was measured with a Thorlabs S145C integrating sphere InGaAs photodiode energy meter. The laser pulse energy was adjusted through a range from ~0.1 pJ up to ~3 pJ. Using the photodiode responsivity curve and the integrator calibration, one can plot measured energy using the photodiode/integrator vs. measured energy using the Thorlabs energy meters. The photodiode response curve is an average of 100 pulses. The detector is now ready for commissioning in the FAST beam line.

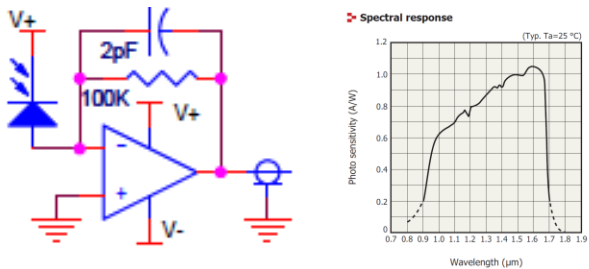

Figure 5: Schematic of the (left) amplifier circuit and (right) spectral response curve of the Hamamatsu InGaAs photodiode G11193-10R.

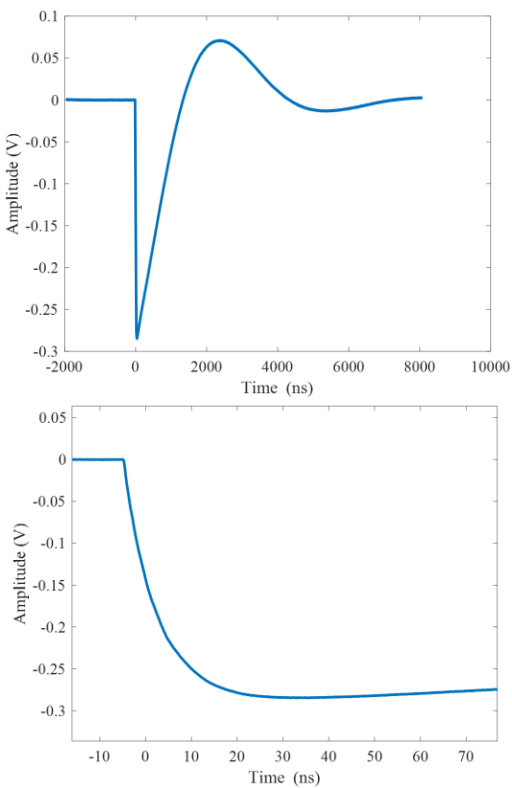

Figure 6: The amplifier response to a single ~3-ps rms calibration pulse from a 1550-nm Erbium fiber laser on a long (top panel) and a short (bottom panel) time scale. The maximum amplitude (~-0.3 V), corresponding to the integrated radiation energy, was measured as a function of the input radiation pulse energy.

Figure 6 shows an example response of the amplifier to a ps-long calibration pulse over two different time scales. Calibration pulses at two wavelength (1054 nm and 1550 nm) were used to calibrate the amplifier's response to single pulses 0.1 – 3 pJ in energy. The results of this calibration are presented in Figure 7. The difference in calibrations for two different wavelengths is being investigated and might be related to the Si photodiode energy meter at 1054 nm, which is close to the limit of its sensitivity range.

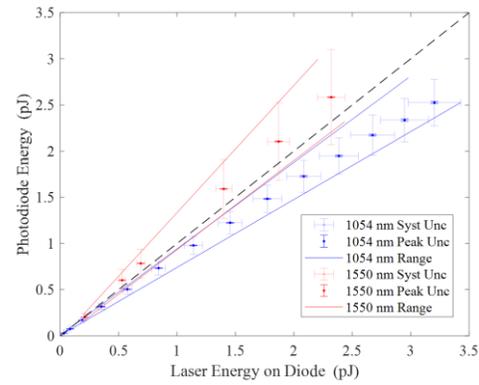

Figure 7: Results of the photodiode calibrations with single laser pulses.

## SUMMARY

A single-pulse transition radiation spectrometer has been developed for the wavelength range of 0.9 – 2 μm with samplings by a series of selectable 100-nm-wide IR bandpass filters. The expected pulse energy is about 1 pJ for a signal corresponding to a shot-noise density fluctuations level in the electron bunch. Actual beam measurements are planned at FAST later this year. Our Year-1 goal is to carry out these measurements with a single-bunch, 0.1-2 nC charge at 45 MeV.

## ACKNOWLEDGEMENTS

This project is supported by a grant (DE-SC0022196) from the DOE ARDAP office.

This research is also supported by the University of Chicago and the US Department of Energy under contracts DE-AC02-76SF00515 and DE-AC02-06CH11357.